\documentclass[onecolumn,times,12pt]{article}
\usepackage{amsfonts,amsmath,amsthm,amssymb,mathrsfs,latexsym,cancel}
\usepackage[dvips]{graphicx}
\usepackage{booktabs,fancyvrb,relsize}
\usepackage{fancyvrb,relsize,hyperref}
\usepackage{chngpage}
\usepackage[square,numbers,round,comma,sort&compress,longnamesfirst]{natbib}

\title{Fundamental and subharmonic transition to turbulence in zero-pressure-gradient flat-plate boundary layers}
\author{Taraneh Sayadi, Curtis W.\ Hamman and Parviz Moin \vspace{6pt} \\ Center for Turbulence Research, \\ Stanford University, Stanford, CA 94305, USA}
\begin{document}

\maketitle

\begin{abstract}
In this fluid dynamics video, recent simulations of transition to turbulence in compressible ($M_\infty = 0.2$), zero-pressure-gradient flat-plate boundary layers triggered by fundamental (Klebanoff K-type) and subharmonic (Herbert H-type) secondary instabilities of Tollmien-Schlichting waves are highlighted.
\end{abstract}

\section{Motivation and Objectives}
Transition in wall-bounded flows is sensitive to the type of disturbance \cite{schlatter10b}.  In this fluid dynamics video, we examine recently obtained direct numerical simulation databases of two boundary layers from laminar (Blasius $\textrm{Re}_\theta = 210$) inflow that transitions into fully turbulent flow via two different physical disturbance transition scenarios:
\begin{enumerate}
\item Fundamental (Klebanoff K-Type) transition  ($\textrm{Re}_{\theta_\textrm{max}} = 1410$) 
\item Subharmonic (Herbert H-Type) transition  ($\textrm{Re}_{\theta_\textrm{max}} = 1250$)
\end{enumerate}
These simulations were conducted to evaluate effects of physically different inflow conditions upon the establishment of organized coherent motions far downstream of the original Tollmien-Schlichting disturbance and compare with the ``forest of hairpins'' \cite{wumoinaps} found in the fully turbulent region of zero-pressure-gradient turbulent boundary layers forced by free-stream turbulence that triggers bypass transition upstream.  For the non-bypass K-Type and H-Type transition scenarios, we report that no forest of hairpins, i.e.\ a highly organized array of complete hairpin vortices identified by isocontours of Q-criterion, is found in the fully turbulent flow.  We propose a possible explanation for this structural discrepancy based upon the different mechanics of these transition scenarios, but do find that near-wall mean statistics of K/H-type transition in the fully turbulent flow are quantitatively similar to that seen in bypass transition \cite{sayadi}.

\section{Video Description}
This fluid dynamics video shows simulations of fundamental and subharmonic transition to fully turbulent flow across a compressible ($M_\infty = 0.2$), zero-pressure-gradient flat plate boundary layer.  For each type of transition (fundamental K-Type and subharmonic H-type), we show one clip from an instantaneous snapshot of the flow and another clip focused on a particular streamwise location ($\textrm{Re}_\theta = 478$ for K-type, $\textrm{Re}_\theta = 618$ for H-type) where the skin-friction transitions from the laminar to turbulent values.  Fixed on that location, several successive snapshots in time of the flow are shown and development of coherent motions made visible by isosurfaces of Q-criterion colored by streamwise velocity.  The H-type transition also has a more pronounced overshoot in skin-friction relative to K-type transition.  In the skin-friction plot, the turbulent correlation (upper dashed purple line) corresponds to $C_f = (0.455)/\log^2(0.05\textrm{Re}_x)$ while, for the other (laminar) profile, $C_f = 0.664/\sqrt{x}$.

Three spanwise numerical rakes (the red, blue and green cylinders in the middle of the image) record the instantaneous streamwise velocity signals shown in the upper-left plot (with matching red, blue and green colors).  The wall-normal position of each rake is at a fixed height (corresponding to $y^+ \approx 5$, $y^+ \approx 50$ and $y^+ \approx 300$ at the end of the domain).  The streamwise position of the rakes is also marked by a red line in the lower-right plot of the time and spanwise-averaged skin-friction profiles along the plate.  The red dot tracks the spanwise-averaged skin-friction profile at the current rake position.  The linear growth of the waves in the initial part of the domain are not shown explicitly, but are visible in the spanwise-averaged skin-friction profiles as traveling waves that grow and decay.  This is clearly seen in the spanwise-averaged skin-friction profiles that travel along the plate during the time-evolving clips.  Additionally, the oscillation of the turbulent spot in the K-type transition follows the phase of the leading TS wave as seen in the skin-friction plot, which also appears as spikes in the streamwise velocity signals from the spanwise rakes.  At the end of these four clips, instantaneous snapshots of the Q-criterion isosurfaces for the both the H/K-type transition are shown at successive downstream locations; the style and viewing angle was chosen to conform with that used by \citet{wumoinaps} in the 2009 APS Gallery of Fluid Motion.  The archived numerical simulation database used to make this video is available at the \href{http://www.stanford.edu/group/ctr/}{Center for Turbulence Research}.

\section{Physical Description}
Disturbances are introduced inside a laminar boundary layer via a numerical disturbance strip. This process represents the receptivity and response of a boundary layer to either external or internal perturbations (see \citet{kachanov94} for a comprehensive review).   One consequence of the receptivity process is the formation of low-amplitude, two-dimensional and unsteady Tollmien-Schlichting (TS) waves.  These two-dimensional TS waves grow exponentially and, when they reach a certain amplitude (near $1$ to $2 \%$ of the free stream velocity), the flow becomes highly three-dimensional eventually leading to fully turbulent flow downstream.  \citet{klebanoff62} and \citet{kachanov84} also studied this type of transition experimentally and showed that in this transitional scenario the two-dimensional TS wave interacts with a set of oblique waves with the same frequency as the fundamental wave, which permits vortex stretching.  Fundamental transition of this kind is also named after Klebanoff as K-type transition.

This non-linear interaction causes the growth of higher harmonics and leads to the formation of $\Lambda$ vortices. These vortices are aligned and cause the flow to lift up at the tip of the vortex and create a region of high shear.  Experimental observations deduced this from the sudden appearance of ``spikes'' in oscilloscope traces of the streamwise velocity signal.  The $\Lambda$ vortices evolve and form hairpin loops that eventually create turbulent spots leading to turbulence.  Oscilloscope ``spikes'', non-staggered $\Lambda$ vortices, and turbulent spots are each visualized in this fluid dynamics video.  In particular, the two spanwise rakes nearest the wall show this lift up process and associated velocity spikes in the profile (four of them at a given time since the spanwise domain length is four times the fundamental spanwise wavelength associated with these flow structures).

An alternative non-linear wave process that leads to transition from laminar to turbulent flow was studied by \citet{kachanov84}.  This transition scenario is initiated from low-amplitude waves inside the boundary layer, and in this respect is similar to K-type transition. However, in this case no turbulent spots, as seen in controlled K-type transition, are observed.  Moreover, the $\Lambda$ vortices are formed in staggered arrangements in contrast to the aligned rows seen in the K-type transition. It was later explained by \citet{herbert88} that, in this transition process, the two-dimensional TS wave interacts with a set of oblique waves with half the frequency of the fundamental wave, i.e.\ the subharmonic wave, resulting in a staggered arrangement of the $\Lambda$ vortices with twice the streamwise length. This subharmonic transition is also named Herbert as H-type transition.  These coherent $\Lambda$ vortices are visualized in this fluid dynamics video and and seen to evolve into hairpin loops preceding the initiation of turbulence and filling of the full turbulent energy spectrum in the velocity signals.  In this subharmonic transition the harmonics of the fundamental wave grow very weakly (leading to a more delayed transition relative to the K-type transition), however, once the subharmonic disturbances grow strong enough, three-dimensional quasi-random disturbances develop causing vortex stretching and eventual breakdown.

These transition scenarios (K/H-type) belong to the class of controlled transitions. The receptivity process is controlled such that it leads to generation of prescribed two- and three-dimensional waves with low amplitudes. These disturbances grow as predicted by linear stability theory and during this process no external nor internal perturbations are allowed to interact with the boundary layer unlike the continuous perturbations applied to the near-wall boundary layer when strong free-stream disturbances are present.

Free stream turbulence (FST) is a form of external perturbation which can cause the boundary layer to transition. If the turbulence intensity is high enough (order of $1\%$ or more), the exponential growth of TS waves is bypassed and transition occurs rapidly and at lower Reynolds numbers than reported for the H/K-type transitional regimes. This type of transition is known as bypass transition. The transition process is initiated by imposing the external FST over a laminar boundary layer flow. Through the receptivity process, disturbances are generated within the boundary layer which lead to streaky structures in stream-wise velocity, this is followed by regions of localized turbulence (turbulent spots) that appear randomly throughout this section and, ultimately, the boundary layer becomes fully turbulent.  In a recent calculation by \citet{wu10}, bypass transition was initiated by periodically feeding an isotropic turbulent box into the inflow and allowed to convect the entire length of the boundary layer in the freestream. It was demonstrated that through this receptivity process, $\Lambda$-shaped vortices are formed inside the boundary layer. Each $\Lambda$-shaped vortex in turn evolves into a packet of three hairpin vortices. These vortices appear randomly inside the transitional regime unlike the ordered arrangement seen in controlled H/K-type transition.

The key difference between the current two controlled transition scenarios and the turbulence that results from free-stream disturbances applied by \citet{wu09} is that no ``forest of hairpins'' is found far downstream (at higher Reynolds numbers) in the H/K-type transition scenarios.  The memory of these upstream disturbances fades away as non-modal interactions, which were originally very small but random perturbations to the disturbance strip forcing, become significant.  Spanwise energy spectra of the streamwise velocity, which can be computed by Fourier transforms from the velocity signals recorded by the three rakes in the fluid dynamics video, show this growth of the modes that are not integer multiples of the fundamental (or subharmonic) forcing.  As a result, the hairpin motions become less organized in the late-transitional regime whereupon a fully populated spanwise turbulent energy spectrum is established with no preference for modal interactions.

Free-stream disturbances, on the other hand, are found to sustain a forest of hairpins \cite{wumoinaps}.  FST persists far downstream providing a means to excite the same fundamental modes responsible for hairpin motions far into the fully turbulent flow.  As a result, FST triggers these same mechanisms further downstream maintaining coherent hairpin packets whereas the convective instability in H/K-type is largely non-existent far downstream.  In spite of these differences, the mean near-wall statistics are quantitatively the same so that, for modeling purposes, incipient hairpin dynamics in near-wall turbulence has merit.  The mean near-wall turbulent statistics from DNS of fundamental, subharmonic and FST disturbances data are found to compare well with physical experiment and each other.  As a result, the vortex regeneration mechanisms associated with hairpin-type coherent structures play a prominent role in near-wall turbulence dynamics and are able to provide a physical foundation for modeling wall-bounded turbulent shear flows.

\section{Simulation Setup}
The code used to carry out these calculations is a compressible Navier-Stokes solver.  The flow has a freestream Mach number $M_\infty = 0.2$.  The fluid is simulated as an ideal gas and Sutherland's law for viscosity is employed. Conservation of kinetic energy in the inviscid limit is achieved by staggered collocation of velocities with respect to density and total energy. A structured, curvilinear coordinate system allows the use of fourth-order spatial derivatives in all three directions.  All convective terms in the Navier-stokes equation are treated implicitly in the near wall region. An implicit A-stable second order scheme is also applied to all viscous terms in wall normal direction close to the wall. All the remaining terms in the Navier-Stokes equation are advanced explicitly in time using an RK3 scheme. Scalable parallelization is achieved using message passing interface (MPI).  Simulations were performed on both the BlueGene/L system at Lawrence Livermore National Laboratory and the BlueGene/P system at the Argonne Leadership Computing Facility on 32k cores for most production runs.
\begin{table}
\begin{adjustwidth}{-0.65in}{0.65in}
\begin{tabular}{cccccccccccc}
\toprule
                &  $L_x$ &$L_y$ & $L_z$ & $N_x$ & $N_y$ & $N_z$ &$\Delta x^+$ &$\Delta y^+_{min}$ &$\Delta z^+$ &$\textrm{Re}_{\theta_\textrm{max}}$ & $\lambda_z$\\
\midrule
H-type    &   9.6      &  1.0    & 0.605  &  4096   & 550      & 512      & 10.684         &0.414            & 6.720             & 1250                        &  0.1514\\
K-type    &   8.599 &  0.92  & 0.606   &  4096   & 550      & 512      & 10.156        &0.405            & 5.724             &1410                         & 0.1515\\
\bottomrule
\end{tabular}
\end{adjustwidth}
\caption{Characteristics of the different grids used for each case.}
\label{tab:grid}
\end{table}

The inlet Reynolds number based on the distance from the leading edge is $Re_x = 10^5$, and the characteristic length scale used for non-dimensionalization is based on the distance of the inlet station from the leading edge of the plate $x_0 = 1$.  At the inflow, the Blasius boundary is $\textrm{Re}_\theta = 210$ in both simulations. In the case of the H-type transition the dimensions of the computational domain are $1 \le x/x_0 \le 10.6$ in the stream-wise direction, $0 \le y/x_0 \le 1$ in wall normal direction and $0 \le z/x_0 \le 0.6055$ in the span wise direction. This span-wise length is $5.2\delta_{99}$, where $\delta_{99}$ is the boundary layer thickness at $\textrm{Re}_\theta = 1250$ close to the end of computational domain.  Spanwise two-point correlations not shown here also confirm that the width of the domain is adequate. The spanwise domain length is also four times the wavelength of the unstable oblique wave introduced into the computational domain at the disturbance strip. The freestream velocity is fixed so that the freestream Mach number $Ma_\infty = 0.2$, non-dimensionalized by the local sound speed.

\section{Summary}
Transition to turbulence via spatially evolving secondary instabilities in compressible ($M_\infty = 0.2$), zero-pressure-gradient flat-plate boundary layers is numerically simulated for both the Klebanoff K-type and Herbert H-type disturbances and visualized in this fluid dynamics video. The objective of this work is to evaluate the universality of the breakdown process and coherent motion development between different routes through transition in wall-bounded shear flows \cite{sayadi}. Each localized linear disturbance is amplified through weak non-linear instability that grows into $\Lambda$-vortices and then hairpin-shaped eddies with harmonic wavelength, which become less organized in the late-transitional regime once a fully populated spanwise turbulent energy spectrum is established.  No forest of hairpins is found far downstream.  Memory of these upstream disturbances fades away as non-modal interactions become significant and multiples of the fundamental (or subharmonic) wavelength are not preferentially seen in the streamwise velocity profiles (as shown in the video), but still remain dynamically significant even though they do not readily stand out above the fray unless preferentially perturbed (e.g.\ as free-stream turbulence perturbs the near-wall layer).

In contrast, free-stream disturbances are able to sustain a forest of hairpins \cite{wu09,wu10,wumoinaps}.  The key physical difference between the K/H-type transition and bypass transition produced by free-stream turbulence is that the FST persists far downstream providing a means to excite the same fundamental modes responsible for hairpin motion growth and development far into the fully turbulent flow.  In effect, FST in such simulations triggers these same mechanisms further downstream maintaining coherent hairpin packets whereas the convective instability in H/K-type is largely non-existent far downstream.  Despite these differences, the mean statistics are quantitatively similar.  As a result, incipient hairpin dynamics in near-wall turbulence has merit as a physical modeling tool at higher Reynolds numbers (appearing directly in flows disturbances, such as FST, perturb the near-wall layer exciting the fundamental modes responsible for hairpin generation in the transitional regime).  The mean near-wall turbulent statistics from DNS of fundamental, subharmonic and FST disturbances data are found to compare well with physical experiment and each other.  Hence, the vortex regeneration mechanisms associated with hairpin-type coherent structures play a prominent role in near-wall turbulence dynamics and provide a foundation for modeling wall-bounded turbulent shear flows.

~\\
{\footnotesize \noindent {\bf Acknowledgments.}\, Computational support provided by the Argonne Leadership Computing Facility at Argonne National Laboratory, supported by the Office of Science of the DOE, contract DE-AC02-06CH11357; DOE National Nuclear Security Administration, award number DE-FC52-08NA28613; and MRI-R2: Acquisition of a Hybrid CPU/GPU and Visualization Cluster for Multidisciplinary Studies in Transport Physics with Uncertainty Quantification, award number 0960306 funded under the American Recovery and Reinvestment Act of 2009 (Public Law 111-5).  C.W.\ Hamman acknowledges support from a DOE Computational Science Graduate Fellowship, grant number DE-FG02-97ER25308.}
\vspace{-0.425in}
\bibliographystyle{unsrtnat}
\bibliography{main}
\end{document}